\documentclass[%
preprint,
 amsmath,amssymb,
 aps,
]{revtex4-1}

\usepackage{graphicx}
\usepackage{dcolumn}
\usepackage{bm}

\begin{document}


\title{Quantum Square Well Bound States Described by Equations with Non-locality}

\author{Toru Ohira}
  \email{ohria@math.nagoya-u.ac.jp}
\affiliation{%
Graduate School of Mathematics, Nagoya University, Nagoya, Japan
}%


\date{\today}

\begin{abstract}
 We present here a simple equation explicitly incorporating non-locality, which reproduces quantized energy levels of the bound states for the square well potentials. 
Introduction of this equation is motivated by studies of differential equation with time 
delayed feedback, which can be viewed as describing temporal non-locality.
\end{abstract}

\pacs{03.65.-w,02.30.Ks}
\maketitle


``Stochasticity'' and ``non-locality'' are the two main elements, which separate quantum mechanics from classical mechanics. Probabilistic dynamics governed by the Schr\"{o}dinger equation and non-local correlations as manifested by the Bell's inequalities are representative features of quantum theory\cite{sakurai}. 

Due to these factors of stochasticity and non-locality, quantum mechanics has attracted much attention in the light of measurement theories\cite{bell,wheeler}. It appears that puzzles still remain (e.g.\cite{pearle,pusey}), and explorations, such as quantum computations and quantum information theories, are actively pursued\cite{mermin}. 

There have been some investigations to reformulate quantum and quantum field theories with respect to the former factor of stochasticity. The main approach is to include a ``noise'' term explicitly in the formulation\cite{nelson,parisi}. However, to the author's knowledge, analogous attempts regarding the latter factor of non-locality have not been made. 

Against this background, we here focus on non-locality and present a simple equation, which takes this factor at its ``face value''. In concrete, we present a simple equation incorporating non-locality, which gives discrete energy spectrums for quantum square well potentials. The presentation of this equation is motivated by considerations of non-locality on the time axes, which are often described by ``delay differential equations''. These equations are often used to describe dynamics under the influence of time delays as in the physiological feedbacks. Though less known in physical context, applications are quite wide including reproductions of blood cells, human posture balancing, neural networks, market dynamics and so on(e.g.\cite{tamas}). Also, mathematically, they have been of interest as they produce rather intricate and complex dynamics to otherwise simple systems just by an increase of the value of the delay parameter\cite{mackey}.

The equation we present in the following is the simplest first order equation of this kind with
re--interpretation of the temporal non-locality of delay as spatial non-locality. With a quantization rule of imposing oscillating dynamics inside the potential well, we show that it can reproduce quantized energy levels as given by the standard procedures of quantum mechanics\cite{schiff,french}.

Let us start describing our equations. The first one is for the 
square well with infinite boundaries, as shown in Figure 1(A). 
The equation is simply given as follows.
\begin{equation}
{d \mu(x) \over d x} = \begin{cases}
                    {(i)^{1+p}} k \mu(x-{L \over 2}), & (0 \leq x \leq {L \over 2})\\
                    {(i)^{1+p}} k \mu(x+{L \over 2}), & (-{L \over 2} \leq x \leq 0)\\
                    0, & ( {L \over 2} < \lvert x \rvert),
                    \end{cases}
\end{equation}
where $i = \sqrt{-1}$, $k = {\sqrt{2 m E} \over \hbar}$ with mass, $m$, and
energy $E$ of the quantum particle, and  $\hbar = {h \over {2 \pi}}$, $h$ being the Plank's constant. $p$ is a parameter which takes values $0, 1$.),
The quantization condition is imposing a condition that, within the well, the
function $\mu(x)$ admits only the oscillatory form. Namely, 
\begin{equation}
\mu(x) \sim e^{i \omega x}
\end{equation}
For $p=0$, substituting this into the equation (1) yields,
\begin{equation}
i \omega = i k \cos({{\omega L} \over 2}), \quad
0 = k \sin({{\omega L} \over 2}).
\end{equation}
These together leads to a quantization, $\omega^2 = k^2$ and
$k_n = {{2 n \pi} \over L}, n = 1, 2, 3, \dots$. 
The associated wave function can be constructed up to the normalization constant as
\begin{equation}
\psi(x) \sim \begin{cases}
               (\mu(x) - \mu^{*}(x))/2 & (-{L \over 2} \leq x \leq {L \over 2})\\
               0 & ( {L \over 2} < \lvert x \rvert )
             \end{cases}
\end{equation}

Similarly, with $p=1$, we obtain the other solution sets with 
$k_n = {{\pi} \over L}(n+1), n = 0,  2, 4, \dots$, with the associated wave
function as
\begin{equation}
\psi(x) \sim \begin{cases}
               (\mu(x) + \mu^{*}(x))/2 & (-{L \over 2} \leq x \leq {L \over 2})\\
               0 & ( {L \over 2} < \lvert x \rvert)
             \end{cases}
\end{equation}
These are well known results of the quantum bound states for this potential\cite{note1}.

\begin{figure}[h]
\begin{center}
\includegraphics[width=0.8\columnwidth]{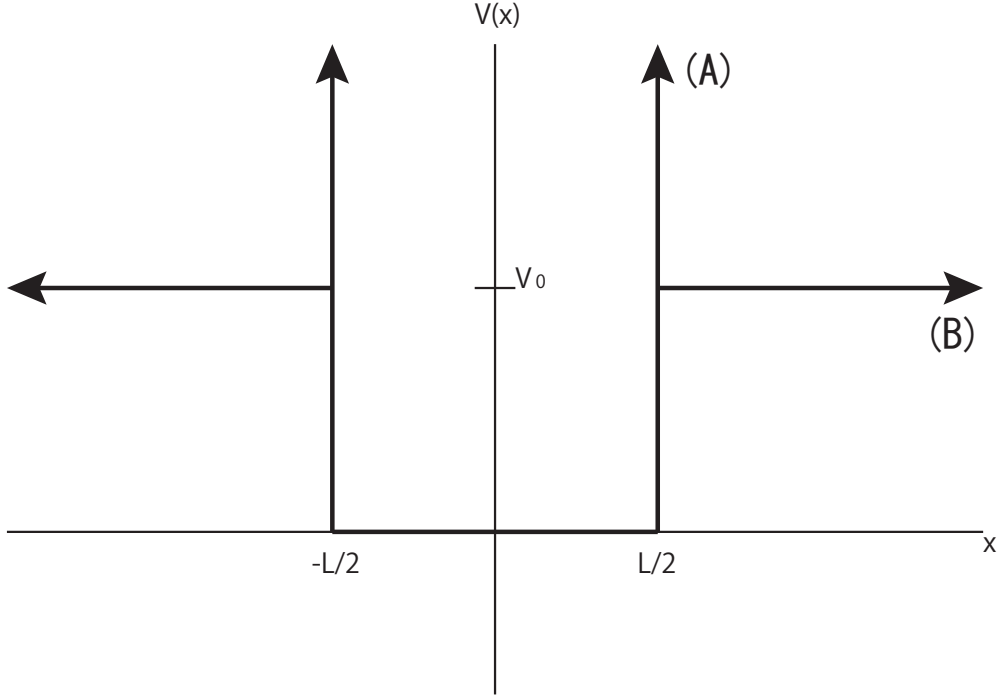}
\end{center}
\caption{Quantum square well with infinite (A) and finite (B) barrier height.}
\label{tree}
\end{figure}

When the hight of potential is finite with $V_0 > E$ (Figure 1(B)), we add
a linear term in the Equation (1).
\begin{equation}
{d \mu(x) \over d x} = \begin{cases}
- \alpha \mu(x) & ({L \over 2} < x )\\
- \alpha \mu(x)  + {(i)^{1+p}} \gamma \mu(x-{L \over 2}) & (0 \leq x \leq {L \over 2})\\
+ \alpha \mu(x) +  {(i)^{1+p}} \gamma \mu(x+{L \over 2}) & (-{L \over 2} \leq x \leq 0 )\\
+ \alpha \mu(x) & ( x< -{L \over 2} ),
\end{cases}
\end{equation}
where $\alpha = {\sqrt{2 m (V_0 - E)} \over \hbar}$ and $\gamma = {\sqrt{2 m V_0} \over \hbar}$. We note an ordinary relation between these parameters, $k^2 + \alpha^2 = \gamma^2$.

By going through the same procedure of imposing the condition of 
Equation (2), we obtain the sets of equations for $p=0,1$.

\noindent
For $p=0$, 
\begin{equation}
i \omega = i \gamma \cos({{\omega L} \over 2}), \quad
\alpha = \gamma \sin({{\omega L} \over 2}),
\end{equation}
leading to $\omega^2 + \alpha^2 = \gamma^2$ and
\begin{equation}
{{\alpha L}\over 2} = {{\omega L}\over 2}\tan({{\omega L}\over 2})
\label{fp1}
\end{equation}
\vspace{1em}

\noindent
For $p=1$, 
\begin{equation}
i \omega = i \gamma \sin({{\omega L} \over 2}), \quad
\alpha = -\gamma \cos({{\omega L} \over 2}),
\end{equation}
leading again to $\omega^2 + \alpha^2 = \gamma^2$, and
\begin{equation}
{{\alpha L}\over 2} = -{{\omega L}\over 2}\cot({{\omega L}\over 2})
\label{fp0}
\end{equation}
By identifying $k = \omega$, Eqs. (\ref{fp1},\ref{fp0} ) give the standard quantum energy levels for this potential.

Also, the associated wave functions can be constructed, for $p=0, 1$,
\begin{equation}
\psi(x) \sim \begin{cases}
               e^{+\alpha x} & (x< -{L \over 2} )\\
               (\mu(x) + {(-1)^p} \mu^{*}(x))/2 & (-{L \over 2} \leq x \leq {L \over 2})\\
               e^{-\alpha x} & ( {L \over 2} < x)
             \end{cases}
\end{equation}

Normally, quantizations with square well potentials are done through physical considerations at the boundaries. Here, in a sense, boundaries of the potential are incorporated into the equation itself as a non-local element, and the quantization 
condition is given by requirements of the oscillatory nature of the solution. It is yet to be investigated that this type of approach can be developed for obtaining or approximating quantum bound states for more general types of potentials.

The proposed equation contain only the first order derivative in space in contrast to the Schr\"{o}dinger equation. In this sense, it appears simpler. However, in general, introduction of non-local term can give rise to quite intricate complex dynamics as shown by investigations of delay differential equations. Even though we have only focused on simple oscillatory solutions, it will be interesting to see whether some of these complex dynamical aspects of general non-local equations play roles in quantum mechanics.

Inversely, from a point of view of studies of delay differential equations, the equation here is new or not well known as it contains imaginary feedback term. Investigations of such types may reveal more interesting aspects of temporal non-locality.

In summary, this report is a modest step to provide an alternate view on some aspects of quantum
mechanics by incorporating non-locality directly into the basic equation. A question whether this style of approach can be further developed to deepen our understanding of quantum mechanics is yet to be explored.

\begin{acknowledgments}
The author would like to thank Philip M. Pearle, Professor Emeritus of Hamilton College, for his insightful comments.
\end{acknowledgments}


\begin{thebibliography}{1}
\bibitem{note1}
Note that the wave function $\psi(x)$ itself does not satisfy the differential equation.
\bibitem{note2}
Note that combinations of the values of $p$ and the parities of the wave functions $\psi(x)$ are different between the
finite and the infinite barrier potentials.
\end{thebibliography}

\end{document}